\documentclass[pra,bibnotes,twocolumn]{revtex4}
\usepackage{graphicx}
\begin{document}
\draft
\def\ds{\displaystyle}
\title{ Topological states at exceptional points}
\author{C. Yuce }
\address{Department of Physics, Eskisehir Technical University, Eskisehir, Turkey }
\email{cyuce@eskisehir.edu.tr}
\date{\today}
\begin{abstract}
We consider an N-level non-Hermitian Hamiltonian with an exceptional point of order N. We define adiabatic equivalence in such systems and explore topological phase. We show that the topological exceptional states appear at the interface of topologically distinct systems. We discuss that topological states appear even in closed systems. We explore dynamical robustness of exceptional edge states.
\end{abstract}
\maketitle

\section{Introduction}

An exceptional point (EP), which appears in non-Hermitian systems determines a phase transition from a real spectrum to a complex spectrum \cite{EP1,EP2,EP3}. An exceptional point of order $n$ is a point in parameter space at which $n$ eigenvalues and the corresponding eigenstates of a non-Hermitian Hamiltonian coalesce. At an EP, a given non-Hermitian Hamiltonian is no longer diagonalizable, but instead can be brought to a matrix containing at least one Jordan block via a similarity transformation. The systems with an EP of order $2$ have been mainly studied in the literature and little attention has been paid to the higher order EPs \cite{EPek1,EPek2,EPek3}. An $N$-level non-Hermitian system can have at most an exceptional point of order $N$ (EPN). It was recently shown that an EPN appears if the non-Hermitian Hamiltonian is nilpotent \cite{underR3}. It is well known that new features inaccessible in Hermitian systems can be seen not only at an EP, but also around an EP. For example, it was shown that two eigenstates of a two state non-Hermitian Hamiltonian are exchanged with each other if the EP is enclosed by a loop in parameter space \cite{dynexcep1,dynexcep2}. An EP is topological in the sense that such an exchange of eigenfunctions occurs even if we deform the loop enclosing the EP. More generally, EPs are believed to play essential roles in the theory of topological insulators in non-Hermitian systems. \\
A promising field of study is the non-Hermitian topological systems. This relatively new research field has recently attracted great deal of attention not only because of theoretical curiosity but also because of unique topological features which may have some technological advantages in the future. It was theoretically shown that topological edge states exist in $1$-D non-Hermitian systems \cite{nonh2,ekl56,PTop3,PTop4}. From the experimental point of view, $1D$ waveguides with loss are generally used to observe topological edge states \cite{sondeney1}. The majority of theoretical works in the context of non-Hermitian topological insulators focus on $1$-D systems such as the complex extensions of Su-Schrieffer-Heeger (SSH) model \cite{1d1,1d2,1d3,1d4,1d5,1d6,majo3,1d7,1d8,1d9,1d10,1d11,1d12,1d13,cemo1,fazla1,fazla2,fazla3,fazla4,fazla5,floquet1,floquet2,floquetekle,fidsak}. Two dimensional systems such as $2$-D photonic graphene with gain and loss \cite{2d1,2d2,2d3,2d4} and a non-Hermitian extension of the Bernevig, Hughes and Zhang (BHZ) model \cite{2d4} have been studied. It was shown that the standard bulk boundary correspondence of topological insulating systems fails in non-Hermitian systems \cite{bulkboun01,bulkboun02,bulkboun03,bulkboun04,bulkboun05,bulkboun06,bulkboun07,bulkboun08,bulkboun09,bulkboun10,bulkboun11,bulkboun12}. The recovery of bulk boundary correspondence under the protection of inversion or inversion-combined symmetries is also demonstrated \cite{,bulkboun07}. The topological transition point of periodical and open systems can be different in non-Hermitian systems in contrast to Hermitian systems. Furthermore, the extension of topological numbers to non-Hermitian systems is not straightforward \cite{winding1,winding2,winding3} and they are generally complex numbers. In \cite{winding4}, a real-valued Berry phase was studied. In a recent work \cite{cempseudo}, the idea of pseudo topological phase was introduced to explain robust topological edge state in a non-Hermitian SSH model. \\
The standard periodic table for Hermitian topological insulators based on the discrete symmetries, particle-hole, time reversal and chiral symmetries, are well known. However, periodic table for non-Hermitian topological insulators have not fully been constructed yet \cite{classification1,classification2}. Bernard-LeClair symmetry classes were proposed for a systematic classifications of non-Hermitian topological insulators \cite{classification1}. Separable, isolated and inseparable bands in the complex energy plane were defined for non-Hermitian periodical systems to study topological phase and construct Chern numbers in 2D separable bands \cite{liangfu}. However, these definitions fail if the system has an EPN for an $N$-level system. In fact, the role of EPs for the topological phase transition has not yet been understood, either. No band gap can be defined in a non-Hermitian system with an EPN and hence topological phase transition has nothing to do with the band gap closing and reopening. In this paper, we explore topological phase in a non-Hermitian topological system with an EPN. We consider an $N$-level non-Hermitian Hamiltonian with an EPN and study topological exceptional states. We define adiabatic equivalence for such systems to study topological phase. We show that a number of interesting features arise in such a system. We explore state conversion and topological states in a closed system. We show for the first time that dynamically robust topological states appear in a non-Hermitian system.

\section{Formalism}

Consider an $N$-level non-Hermitian system. Suppose that the corresponding Hamiltonian $\mathcal{H}$ has an EPN with $E=0$ eigenvalue. An EPN appears if the matrix form of the Hamiltonian is similar to a Jordan block $\mathcal{J}$, where the similarity transformation reads $\ds{\mathcal{J}=S^{-1} \mathcal{H} S}$ and $S$ is a nonsingular matrix. A Jordan block of size $N$ with value $E$ is an $N{\times}N$ square, upper triangular matrix having the value $E$ repeated along the diagonal, all ones along the superdiagonal, and zeros elsewhere. Generally speaking, it is not easy to find a nonsingular $S$ matrix for a given Hamiltonian. Fortunately, there is another way to study such a system \cite{underR3}. A Hamiltonian $\mathcal{H}$ has an EPN with $E=0$ eigenvalue iff it is a nilpotent operator of order $p$, $\mathcal{H}^p=0$, where $p{\leq}N$ is a positive integer.\\
To illustrate our idea, consider a tight binding lattice with asymmetric forward and backward tunneling amplitudes
\begin{equation}\label{65748mu}
\mathcal{H}=  \sum_{n=1}^{N-1} (J^{F}_n~  |n><n+1|+J^{B}_n~|n+1><n|)
\end{equation}
where $J^{F}_n$ and $J^{B}_n$  are site-dependent forward and backward tunneling amplitudes, respectively. The Hamiltonian is not Hermitian unless $\ds{J^{F}_n=(J^{B}_n})^{\star}$. \\
We demand that the matrix form of the above Hamiltonian is similar to a Jordan block of size $N$. Therefore, one may consider unidirectional tunneling amplitude such that either $\ds{J^{F}_n=1, J^{B}_n=0}$ or $\ds{J^{F}_n=0, J^{B}_n=1}$. We refer the reader to the paper \cite{BL9} for a possible physical implementations of a lattice with unidirectional tunneling amplitude. The corresponding two non-Hermitian Hamiltonians are then given by $\ds{\mathcal{H}_1=  \sum_{n=1}^{N-1}    |n><n+1| }$ and $\ds{\mathcal{H}_2=  \sum_{n=1}^{N-1}    |n+1><n|)}$. Their matrix forms are of the form of Jordan block of size $N$ and value $0$. Therefore, they have exceptional points of order $N$. These two special Hamiltonians allow us to study topological phase in a system with an EPN. \\
To study topological insulators, we need topological numbers, which can only be defined for gapped Hamiltonians. As an example of the gapless Hermitian system, consider $J^{F}_n=J^{B}_n=1$. In this case, the system becomes a conductor and hence no topological zero energy edge states appear. We stress that no topological number can be defined for $\ds{\mathcal{H}_1 }$ and  $\ds{\mathcal{H}_2 }$ since they are not gapped Hamiltonians. For example, the non-Hermitian Zak phase for a periodical gapped Hamiltonian is given by $\ds{\gamma_n=\int dk    <\psi_n^L |i\partial_k|  \psi_n^R>  }$, where n labels the band index, $\ds{|  \psi_n^R>  }$, $\ds{|  \psi_n^L>  }$ are the normalized right and left eigenvectors of the Hamiltonian, and the integral is taken over the 1D Brillouin zone. Unfortunately, this formula is ill-defined in our system. One may naturally expect that topological edge states don't appear in the non-Hermitian case, either. Below we will show that the two Hamiltonians $\ds{\mathcal{H}_1}$ and $\ds{\mathcal{H}_2}$ have topological edge states. To do this, we need to define topological phase for a non-Hermitian system with an EPN.\\
We start with the idea that topological properties of a Hamiltonian can not change under the $\it{adiabatic~ deformation}$, which is a fictitious process and does not take place in time. Let us now define adiabatic deformation of a non-Hermitian Hamiltonian with an EPN. Suppose that some parameters of the Hamiltonian are changed continuously. The Hamiltonian is adiabatically deformed if the deformed Hamiltonian remains similar to a Jordan block $\mathcal{J}$ of size $N$ with $E$ during the continuous change of the parameters. In this case, the original and deformed Hamiltonians are said to be adiabatically equivalent. In other words, the topological features of the original system are not changed by such a deformation. The adiabatic equivalence implies that the original and deformed Hamiltonians are assumed to be similar $\ds{ \mathcal{H}=S^{-1} \mathcal{H}^{\prime} S}$, where $S$ is a square nonsingular matrix and $\ds{ \mathcal{H}^{\prime}}$ is the disordered Hamiltonian. Then $\ds{ \mathcal{H}}$ and $\ds{ \mathcal{H}^{\prime}}$ have the same eigenvalue and the exceptional state of $\ds{ \mathcal{H}}$ is protected against the adiabatic deformation of the Hamiltonian.  \\
The most interesting feature of topological systems is the existence of robust topological states. We can now study three interesting cases in our system: i-) Robust exceptional state at the open edges. ii-) Robust exceptional state at the interface of adiabatically inequivalent non-Hermitian systems with EPN. iii-) Robust exceptional state that appears in a closed system. Below, we first study them separately. Then we explore state conversion and dynamical robustness in our non-Hermitian topological system.\\
{\it{ Robust exceptional states at the open edges}}: Suppose that the non-Hermitian lattice has open edges. Since the non-Hermitian Hamiltonian (either $\mathcal{H}_1$ or $\mathcal{H}_2$) is not adiabatically equivalent to vacuum Hamiltonian, zero energy exceptional state appears at the edge of the system. More precisely, the exceptional state is localized at the left edge for the Hamiltonian $\ds{\mathcal{H}_1}$ and at the right edge for the Hamiltonian $\ds{\mathcal{H}_2}$. These asymmetric localizations of the edge states are due to the non-Hermitian skin effect. \\
Let us now study robustness of the zero energy exceptional state. Consider forward (backward) tunneling amplitude disorder for $\ds{\mathcal{H}_1}$ ($\ds{\mathcal{H}_2}$). Therefore, we assume that $\ds{J^{F}_n}$ $\ds{(J^{B}_n)}$ become non-zero, real-valued random numbers for $\ds{\mathcal{H}_1}$  ($\ds{\mathcal{H}_2}$). One can easily construct the matrix forms of the disordered Hamiltonians. It is interesting to note that the matrix forms of the original and disordered Hamiltonians are similar to a Jordan block of size $N$ and value $E=0$. This means that they are adiabatically equivalent. In the presence such disorder, the EPN is not destroyed and the original and disordered systems have the same exceptional eigenstate with zero eigenvalue.\\ 
We perform numerical computation to check our qualitative predictions. Consider the Hamiltonian $\ds{\mathcal{H}_1}$ with $N=10$ sites. We find that the zero energy exceptional eigenstate is given by $\ds{\psi_1=\{  0,0,0,0,0,0,0,0,0,1    \}^T }$. Next we add random  forward tunneling amplitude disorder such that $\ds{J^{F}_{n} }  $ take random values in the interval $\ds{[-1.5,1.5]}$ with the assumption that  $\ds{J^{F}_{n} \neq 0}  $ for any $n$. We numerically see that the eigenstate $\psi_1$ and its zero eigenvalue remain the same. We stress that not only the eigenvalue but also the form of the topological eigenstate resist to the disorder. This is unique to non-Hermitian systems. As it is well known, topological zero energy states in $1D$ Hermitian systems are perturbed in the presence of the disorder to make the eigenvalue remain the same.\\
Let us now consider next-nearest-neighbor (NNN) tunneling amplitudes between the lattice sites. Generally speaking, the additional NNN tunneling energies to the Hamiltonian breaks topological phase. Fortunately, this is not the case in our system. This can be seen from the Hamiltonian $\ds{\mathcal{H}_1}$ with NNN tunneling amplitudes: $\ds{\mathcal{H}_1^{\prime}=  \sum_{n=1}^{N-1}    |n><n+1| +}\sum_{n=1}^{N-2} J^{\prime}   |n><n+2| $, where ${J^{\prime} <}1$ is the NNN tunneling amplitude. This Hamiltonian is similar to the Hamiltonian $\ds{\mathcal{H}_1}$. Therefore, the exceptional state with or without NNN tunneling amplitudes are the same and robust against the tunneling amplitude disorder.
\begin{figure}[t]
\includegraphics[width=8.8cm]{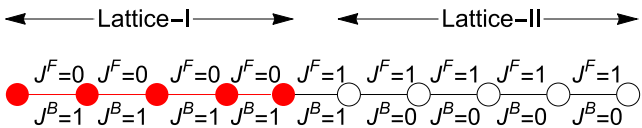}
\caption{The lattice-I on the left and the lattice-II on the right are described by $\ds{\mathcal{H}_1}$ and $\ds{\mathcal{H}_2}$, respectively. Two degenerate topological exceptional states occur at the interface of these two topologically inequivalent systems. They are robust against tunneling amplitude disorder. }
\end{figure}\\
{\it{Robust exceptional states at the interface }}: We stress that the Hamiltonians $\ds{\mathcal{H}_1}$ and $\ds{\mathcal{H}_2}$ are not adiabatically equivalent even if they have EPN. Although they are similar to the Jordan block of size $N$ and value $0$, there is no adiabatic deformation connecting them. To see this, assume that tunneling amplitude are adiabatically changed according to $\ds{J^{F}_n}=\cos(\omega t)$ and $\ds{J^{B}_n}=\sin(\omega t)$, where $\omega<<1$. In this case, the Hamiltonian (\ref{65748mu}) is initially $\ds{\mathcal{H}_1}$ and becomes $\ds{\mathcal{H}_2}$ at ${\omega}t=\pi/2$. The Hamiltonians at intermediate times are not similar to the Jordan block of size $N$ and value $E=0$. For example, the system becomes Hermitian instantaneously at ${\omega}t=\pi/4$ and hence no EPN appears. Consequently, we conclude that $\ds{\mathcal{H}_1}$ and $\ds{\mathcal{H}_2}$ are topologically distinct. According to the standard bulk-boundary correspondence, topological edge states appear at the interface of two topologically distinct systems. Let us study if this statement works in our non-Hermitian system.\\
Consider an interface of the two lattices as shown in the Fig.1. The lattice-I on the left and the lattice-II on the right are governed by $\ds{\mathcal{H}_1}$ and $\ds{\mathcal{H}_2}$, respectively. Suppose that there are $N=10$ sites as shown in the Fig.1. We numerically find two zero energy exceptional states localized around the interface. They are given by $\ds{\psi_1=\{  0,0,0,1,0,-1,0,0,0,0    \}^T }$ and $\ds{\psi_2=\{  0,0,0,0,1,0,-1,0,0,0    \}^T  }$. These two exceptional states are degenerate since they have zero energy. Note that two other eigenstates with $\mp1$ eigenvalues appear in the system, too. Let us now add tunneling amplitude disorder to study robustness of the two zero energy states. Suppose $\ds{J^{B}_n}$ take random values and $\ds{J^{F}_n=0}$ in the lattice-I while $\ds{J^{F}_n}$ take random values and $\ds{J^{B}_n=0}$ in the lattice-II. We numerically see that the degenerate exceptional states resist to such disorder, which is signature of topological feature. To this end, we note that the other two eigenstates with $E=\mp1$ are given by $\ds{\psi_{E=-1}=\{  0,0,0,0,-1,1,0,0,0,0    \}^T }$ and $\ds{\psi_{E=1}=\{  0,0,0,0,0,1,1,0,0,0    \}^T  }$. It is interesting to see that the forms of these eigenstates are not changed although their eigenvalues change with the tunneling amplitude disorder. \\
{\it{Ring lattice}}: Let us now study the system where the two edges of the lattice are connected. In Hermitian $1D$ systems, such systems have no topological states as they have no edges where the topological phase transition occurs. However, this is not always the case in non-Hermitian systems.    \\
Consider the Hamiltonian $\ds{\mathcal{H}_1}$. If the lattice has open edges, then the tunneling amplitudes between the two edges ($n=N$ and $n=1$) are set to zero, $\ds{J^{F}_N}=0$ and $\ds{J^{B}_N}=0$. Suppose now that the two edges of the lattice are connected. Let us choose $\ds{J^{B}_N}=1$ and $\ds{J^{F}_N}=0$. Note that this doesn't lead to the periodic boundary condition, which would be satisfied if $\ds{J^{F}_N}=1$ and $\ds{J^{B}_N}=0$. One can show that the matrix form of the corresponding Hamiltonian is still similar to the Jordan block of size $N$. In fact, the two systems with and without edges are adiabatically equivalent as there exists an adiabatic deformation connecting them. This can be seen by gradually increasing $\ds{J^{B}_N}$  from $0$ to $1$, for which the corresponding Hamiltonian remains nilpotent. Therefore the EPN is not destroyed and the exceptional state survives even if the edges are connected. Furthermore the exceptional state is robust against the same disorder introduced above for the system with open edges. According to the standard theory of topological insulators, topological invariants don't change without a band gap closing. If you move from a topologically nontrivial material into the trivial one, a topological invariant changes, so the gap has to close somewhere along the way. In our system, topological phase transition occurs if the deformed Hamiltonian is not similar to a Jordan block of size $N$. The Hamiltonians with the periodical and open boundary conditions are still similar to each other.\\
{\it{State conversion }}: Consider the non-Hermitian system governed by either $\ds{\mathcal{H}_1}$or $\ds{\mathcal{H}_2}$. In an experiment, one can start with an arbitrary wave packet. However, our system has only one available eigenstate and hence we can not expand the initial wave packet in terms of the eigenstates. No transition between the eigenstates can be defined as in Hermitian systems. A question arises. What is the time evolution of any arbitrary state? Let us study this problem. If the initial state is exactly the exceptional state, then it will preserve its form at any time. If the initial state is not the exceptional state, then we expect that the system will eventually be in the exceptional state since the only eigenstate in the system is the exceptional state. In other words, the state at large times is always the topological state, regardless of the initial state. We perform numerical computation for the Hamiltonian $\ds{\mathcal{H}_1}$ to test this idea. Suppose that the lattice has $N=14$ sites. In the Fig.2, one can see the normalized density plot ($\ds{\frac{|\psi_n|^2}{\sum_n|\psi_n|^2}}$). The single site $n=1$ is initially excited in the Fig.2.(a) ($\ds{\psi_n(0)=\delta_{n,1}}$) and $n=N$ in the Fig.2.(b) ($\ds{\psi_n(0)=\delta_{n,N}}$). In other words, the initial state is prepared in the exceptional eigenstate in (a) while in a state other than the exceptional eigenstate in (b). As can be seen, the former one is dynamically stable while the latter one is not as expected. The wave packet is always localized at $n=1$ site in (a). However, the state in (b) is converted into the exceptional state after a while. We perform numerical computations for various arbitrary initial wave states. We see that they will be eventually the topological exceptional state. We conclude that any state is converted into the topological exceptional state. To this end, we note that the total intensity is constant in (a) but grows in time in (b).
\begin{figure}[t]
\includegraphics[width=3.85cm]{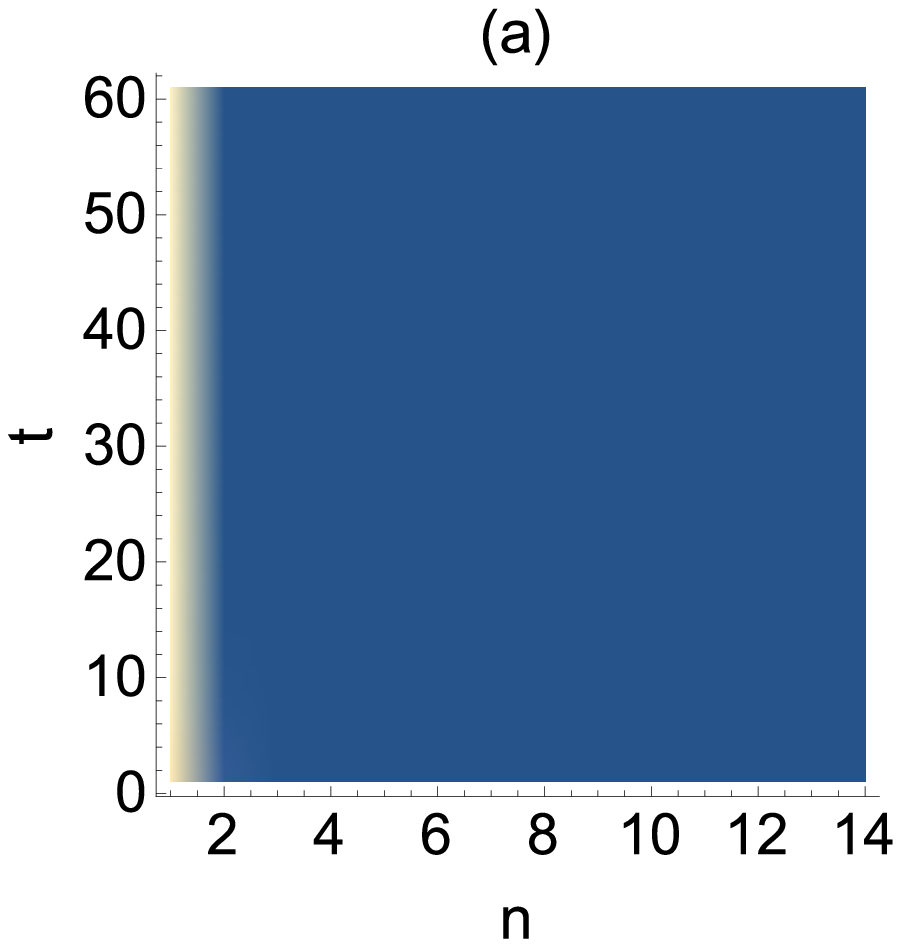}
\includegraphics[width=3.85cm]{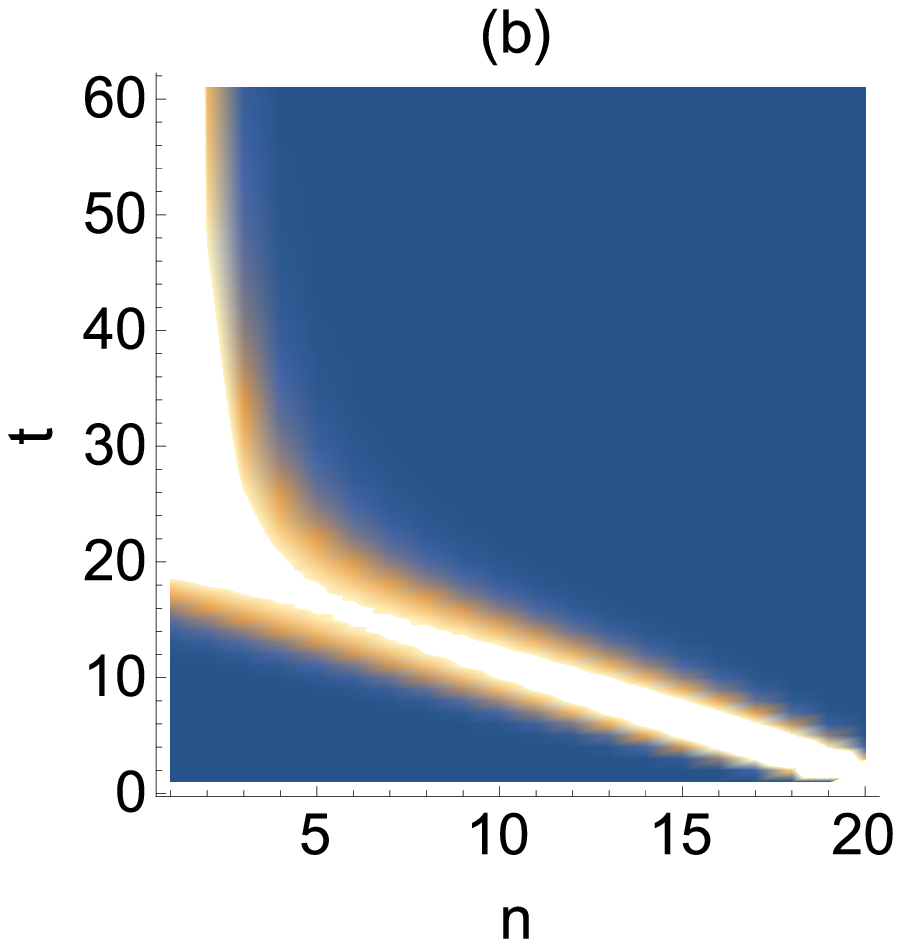}
\caption{The normalized density $\ds{\frac{|\psi_n|^2}{\sum_n|\psi_n|^2}}$ as functions of $n$ and time. The single site $n=1$ is initially excited in (a) ($\ds{\psi_n(0)=\delta_{n,1}}$) and $n=N=14$ in (b) ($\ds{\psi_n(0)=\delta_{n,N}}$). The initial state in (a) is the topological exceptional state and dynamically stable. However, it is not in (b) and it becomes the topological exceptional state after a while. This leads to the transport from the right edge to the left edge. The state conversion is due to the fact that the system has only one available eigenstate.}
\end{figure}
\\
{\it{Dynamical robustness of exceptional edge states}}: The disorder is assumed to be time-independent in the standard theory of topological insulators. In $1D$ Hermitian topological insulators, robustness against time-independent disorders means that the eigenvalues of topological states remain the same provided that the disorder is not so strong to close the band gap. No comprehensive answer has been given in the literature for the question of robustness of topological states subject to time-dependent perturbations \cite{henrik}. \\
We define the dynamical robustness as robustness against time dependent disorder. Let us now explore dynamical robustness of the topological exceptional states. Consider the time-dependent generalization of the Hamiltonian  $\ds{\mathcal{H}_1}$, where the forward tunneling amplitudes $\ds{J^{F}_n}$ are time dependent and $\ds{J^{B}=0}$. In Hermitian systems, time-dependent perturbations leads to transitions among the eigenstates of the system. However, no such transitions occur in our system since there exists only one available eigenstate. This implies that the exceptional state remains robust against the time-dependent forward tunneling. To check the validity of this argument, we solve the equation $\ds{ \mathcal{H} |\psi> =i \partial_t |\psi>   }$. We first assume $\ds{J^{F}_n= (1+0.5 \sin{\omega t})}$ and $\ds{J^{B}_n=0}$ and perform numerical computation for the initial state well localized at the $n=1$ site (the exceptional state). We see that the initial state preserves its form in time as expected. We perform another numerical computation for the parameters $\ds{J^{F}_n= (1+R_n+0.5 \sin{(R^{\prime}_n \omega t)})}$ and $\ds{J^{B}_n=0}$, where $\ds{R_n}$ and $\ds{R^{\prime}_n}$ are independent real valued random numbers in the interval $\ds{[-0.5,0.5]}$. We numerically see that the results are in good agreement with our prediction. The initial zero energy exceptional state preserves its form even when the disorder is time-dependent. We conclude that the system is immune to time-dependent variations of the Hamiltonian provided that there is an EPN in the system. \\
To understand the physics behind the dynamical robustness in our system, let us consider the following two level Hamiltonian with a single Jordan block: $\ds{\mathcal{H}=\left(\begin{array}{ccccc}0   & J(t)  \\ 0 & 0  \end{array}\right)}$ where $\ds{  J(t)   }$ is an arbitrary time-dependent forward tunneling amplitude ($\ds{  J(t)  \neq 0 }$ for all $t$). This Hamiltonian is nilpotent $\ds{\mathcal{H}^2=\textbf{0}  }$ and second order EP occurs. The corresponding solution is given by $\ds{\psi(t) =\left(\begin{array}{cc}1  \\ 0   \end{array}\right)  }$, which does not include time explicitly. We stress that this solution is time-independent for the time-dependent Hamiltonian. This is because of the fact that there is only one state at the EP. It is interesting to see that the time-dependent state is just the instantaneous state. Therefore, the system behaves adiabatically no matter how fast $J(t)$ changes in time. This interesting feature is unique to non-Hermitian system as EPs don't appear in Hermitian systems. 

 \section{Conclusion}
 
The non-Hermitian extension of topological systems is much richer than topological Hermitian systems. Intuitively, one expect that exceptional points play important roles in the non-Hermitian extension of topological systems. In this paper, we have studied robust topological state in a non-Hermitian system with an EPN. We have defined adiabatic deformation of a Hamiltonian with an EPN. We have discussed that topological properties of the Hamiltonian can not change under the adiabatic deformation. Since no band gap is defined in our system, topological phase transition point has nothing to do with band gap closing and reopening. We have shown that the system with an EPN has interesting topological effects. For example, topological states appear in closed $1D$ system, which implies that the standard bulk-boundary correspondence fails in non-Hermitian systems. We have discussed state conversion and found that an arbitrary initial state always evolves to topological exceptional state. We have defined and explored dynamical robustness in our system.

\end{document}